\pdfoutput=1

\documentclass[11pt]{article}

\usepackage[preprint]{acl}

\usepackage{times}
\usepackage{latexsym}

\usepackage[T1]{fontenc}

\usepackage[utf8]{inputenc}

\usepackage{microtype}

\usepackage{inconsolata}

\usepackage{graphicx}

\usepackage{booktabs}
\usepackage{amsmath}
\usepackage{amssymb}
\usepackage{multirow}
	
\usepackage{color, colortbl}
\definecolor{Gray}{gray}{0.9}

\title{MolTRES: Improving Chemical Language Representation Learning for Molecular Property Prediction}

\author{Jun-Hyung Park$^{1}$\quad Yeachan Kim$^{2}$\quad Mingyu Lee$^{2}$\quad Hyuntae Park$^{2}$\quad SangKeun Lee$^{2,3}$ \\
$^{1}$BK21 FOUR R\&E Center for Artificial Intelligence, Korea University \\
$^{2}$Department of Artificial Intelligence, Korea University\\
$^{3}$Department of Computer Science and Engineering, Korea University\\
\texttt{\{irish07, yeachan, decon9201, pht0639, yalphy\}@korea.ac.kr}
}

\begin{document}
\maketitle
\begin{abstract}
Chemical representation learning has gained increasing interest due to the limited availability of supervised data in fields such as drug and materials design. This interest particularly extends to chemical language representation learning, which involves pre-training Transformers on SMILES sequences -- textual descriptors of molecules. Despite its success in molecular property prediction, current practices often lead to overfitting and limited scalability due to early convergence. In this paper, we introduce a novel chemical language representation learning framework, called MolTRES, to address these issues. MolTRES incorporates generator-discriminator training, allowing the model to learn from more challenging examples that require structural understanding. In addition, we enrich molecular representations by transferring knowledge from scientific literature by integrating external materials embedding. Experimental results show that our model outperforms existing state-of-the-art models on popular molecular property prediction tasks.
\end{abstract}

\section{Introduction}
Deep neural networks (DNNs) have emerged as a compelling, computationally efficient approach for predicting molecular properties, with significant implications in material engineering and drug discovery. By training DNNs on molecule data to predict the properties in a supervised manner or to reconstruct molecules in an unsupervised manner, these networks can significantly reduce the costs of traditional methods, which typically require chemical experts and wet-lab experiments. Moreover, DNN-based molecular prediction has gained increasing popularity due to the generalization capacity of DNNs. This allows for the application of a single (pre-)trained model across various tasks, reducing the need for task-specific modeling.

\begin{figure}[t]
\centering
\includegraphics[width = \linewidth]{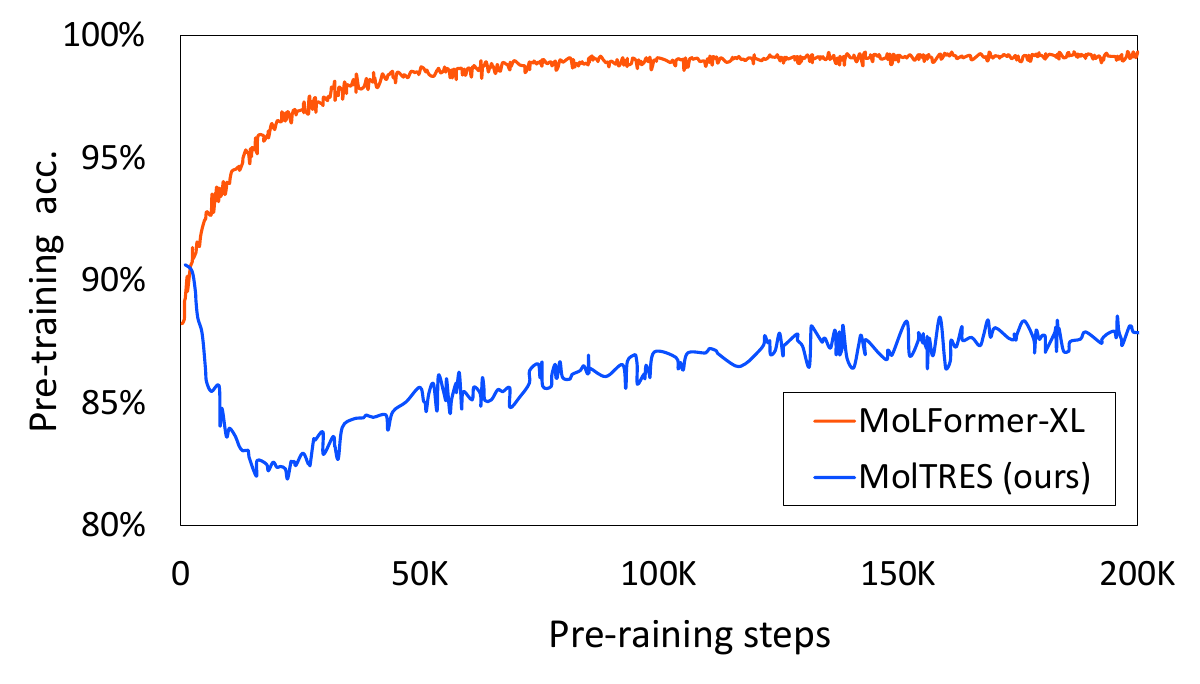}
\includegraphics[width = \linewidth]{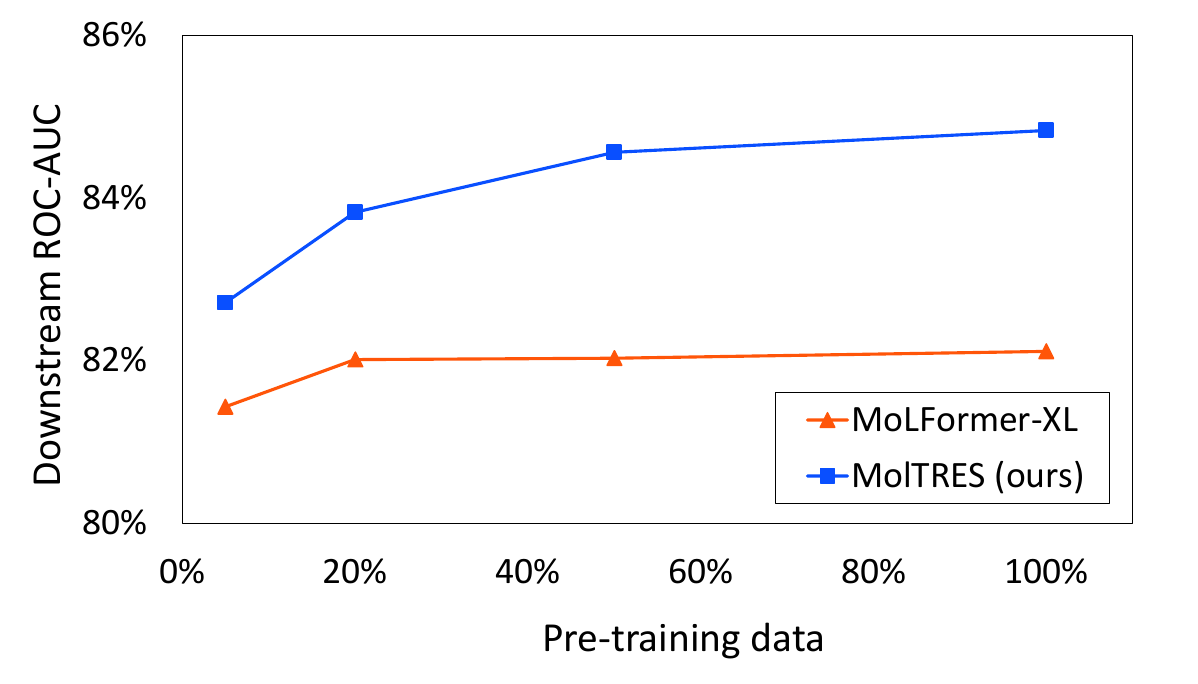}
\caption{Existing pre-training methods for chemical language representation learning already converge at their early stage without seeing the entire data. Consequently, MoLFormer \cite{Ross-NMI2022}, a state-of-the-art chemical language representation learning method, exhibits limited scalability in terms of data size.}
\end{figure}\label{fig:motiv}


Inspired by recent advances in pre-trained language models in the field of natural language processing (NLP), several chemical language representation learning methods based on SMILES Transformers \citep{Wang-BCB2019, Chithrananda-Neuripsworkshop2020} have been proposed. These methods typically employ self-supervised tasks on SMILES (Simplified Molecular-Input Line Entry System) sequences of molecules, analogous to the masked language modeling (MLM) commonly used in BERT \cite{Devlin-naacl2019}. Since modern Transformers are designed to scale to massive NLP corpora \cite{Vaswani-NIPS2017}, they offer practical advantages in terms of efficiency and throughput. This enables the models to leverage massive amounts of SMILES sequences to learn universal representations for molecules, leading to performance improvements in a wide range of molecular property prediction tasks \cite{Ross-NMI2022}. However, as these models typically follow settings designed for natural language modeling, the optimal pre-training settings for chemical language representation learning remain underexplored.

Through extensive investigation into the pre-training of SMILES Transformers, we have discovered that the current pre-training task, MLM on SMILES sequences using a random masking strategy, is not effective for learning informative molecular representations. We have empirically observed that this task can be easily solved using surface patterns, leading to overfitting and limited scalability, as shown in Figure \ref{fig:motiv}. This may be attributed to two inherent properties of SMILES. First, existing large-scale molecule datasets exhibit unbalanced atom distributions \citep{He-ICLR2023}. For example, in ZINC \citep{Irwin-ChemInf2012}, a representative dataset containing billions of molecules, carbon (C), nitrogen (N), and oxygen (O) comprise 95\% of the tokens in total SMILES sequences. Second, the SMILES grammar contains many superficial patterns, such as numbers representing ring structures that always appear twice. These patterns allow the model to predict original tokens without learning the underlying chemical information. Furthermore, unlike natural language, which is fundamentally grounded in concepts and possesses general expressivity across various problem-solving scenarios, SMILES is designed solely to express molecular structure and does not directly represent molecular properties. Thus, the current pre-training task likely provides a limited notion of molecular properties.

In this paper, we propose a novel framework for pre-training SMILES transformers, called MolTRES (\textbf{Mol}ecular \textbf{TR}ansformer with \textbf{E}nhanced \textbf{S}elf-supervised learning), to address the aforementioned issues. Our framework focuses on two key objectives: (1) increasing the difficulty of the pre-training task, and (2) incorporating external knowledge related to molecular properties into model representations. To achieve these goals, we first present a novel dynamic molecule modeling method, coined DynaMol, based on generator-discriminator training \cite{Clark-ICLR2020}. This method trains a model to distinguish real SMILES tokens from synthetically generated replacements, jointly used with substructure-level masking. It facilitates to significantly increase the masking ratio for more challenging training examples, while minimizing discrepancy caused by mask tokens. In addition, we enhance model representations by integrating mat2vec word representations \citep{Tshitoyan-Nature2019} trained on massive scientific literature. This integration helps to directly embody molecular properties in the learned representations.

To demonstrate the effectiveness of MolTRES, we conduct extensive experiments and ablation studies on diverse molecular property prediction tasks. We evaluate MolTRES on eight classification and four regression tasks from MoleculeNet, covering quantum mechanical, physical, biophysical, and physiological properties of chemicals. Our results indicate that MolTRES outperforms state-of-the-art baselines across most tasks, including 1D sequence-, 2D graph-, and 3D geometry-based chemical models. Further analysis shows that MolTRES significantly improves the capabilities of chemical language representation learning by addressing the limitations of existing approaches. Our contributions are summarized as follows:
\begin{itemize}
    \item We propose MolTRES, a novel framework to pre-train SMILES Transformers based on generator-discriminator training and external knowledge transfer.
    \item We present a novel architecture for SMILES transformers efficiently integrated with word representations trained on scientific literature.
    \item Experimental results demonstrate that MolTRES establishes state-of-the-art results over a wide range of molecular property prediction tasks.
\end{itemize}

\begin{figure*}
\centering
  \includegraphics[width=\textwidth]{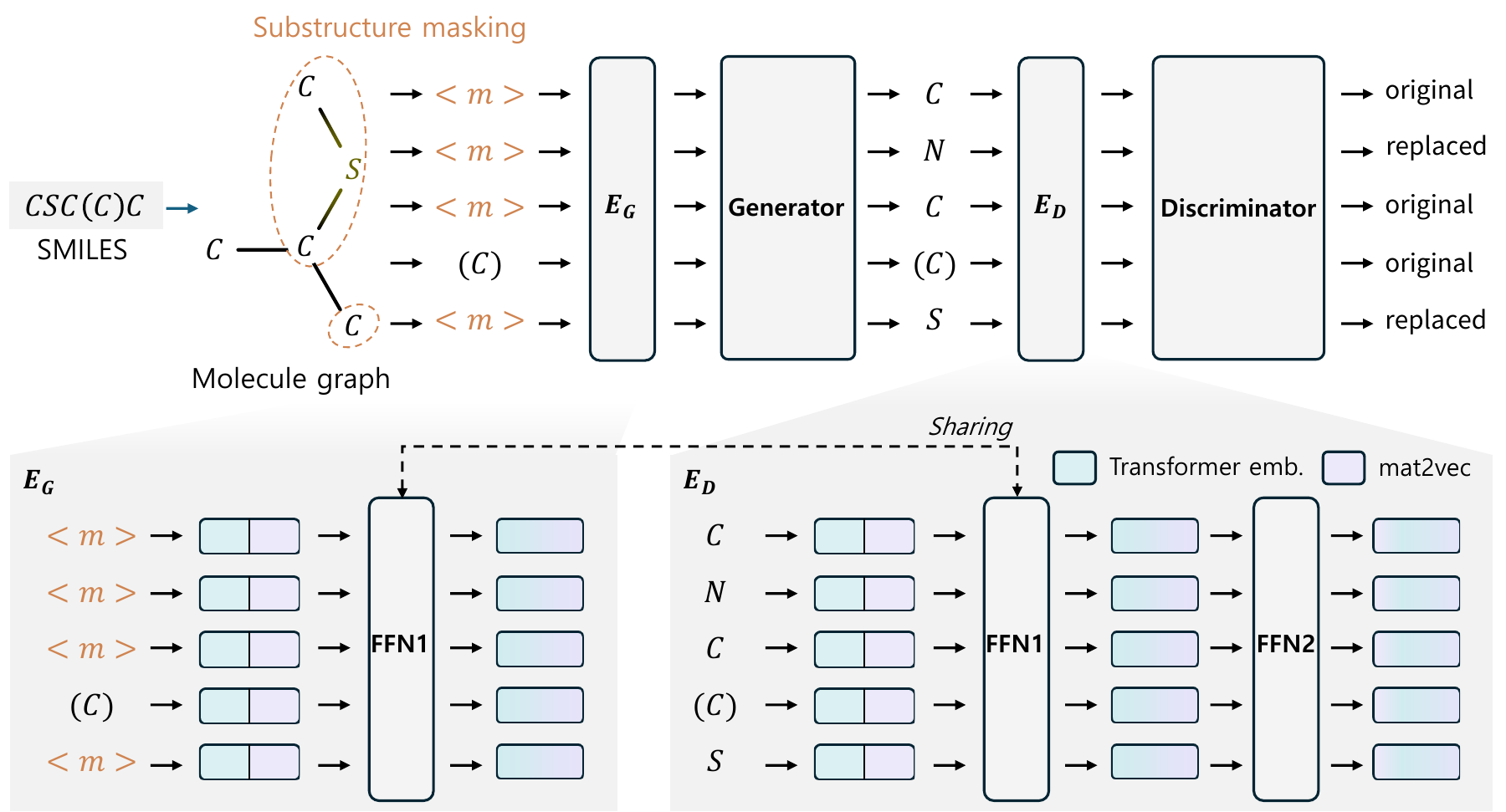}
    \caption{Overview of MolTRES. \textit{E\textsubscript{G}} and \textit{E\textsubscript{D}} represent the embedding layers of the generator and discriminator, respectively. It is noteworthy that the mat2vec embeddings are frozen during pre-training.}
    \label{fig:num_of_retune_prompt}
\end{figure*}

\section{Preliminaries}
\subsection{SMILES Transformer}
Transformer \citep{Vaswani-NIPS2017} is a popular neural network architecture for processing texts, which can also be applied to processing SMILES sequences. It consists of a series of Transformer blocks, each involving a multi-head self-attention layer followed by a multi-layer feed-forward network. The self-attention layer allows the network to effectively model global dependencies within the sequence of input tokens. Given a series of vector representations for tokens, a self-attention layer applies three linear transformations to generate query ($q$), key ($k$), and value ($v$) representations, respectively. The outputs at position $m$ are calculated by aggregating the representations from other positions using a weighted sum, where weights are derived from a similarity function between $q$ and $k$, as follows:
\begin{equation}
    \text{Att}(q, k, v)_{m} = \frac{\Sigma_{n=1}^{N}\text{sim}(q_m, k_n)v_n}{\Sigma_{n=1}^{N}\text{sim}(q_m, k_n)}
\end{equation}
where $\text{sim}(q_m, k_n) = \text{exp}(q_{m}^{\mathsf{T}}k_n/\sqrt{d})$ and $N$ is the length of tokens. Transformer can effectively capture the dependencies in variable-length sequences, and therefore, it is utilized in processing SMILES, as in ChemBERTa \cite{Chithrananda-Neuripsworkshop2020}. However, self-attention shows a quadratic complexity $O(N^2)$ dereived from the computation of the inner product between every token pair, which incurs significant costs when processing molecules represented in long SMILES sequences like polymers. To reduce the complexity,  MoLFormer \cite{Ross-NMI2022} has introduced linear attention with rotary embeddings. This reformulates the original self-attention layer as follows:
\begin{equation}
    \text{Att}(q, k, v)_{m} = \frac{\Sigma_{n=1}^{N} \phi(R_{m}q_m)^{\mathsf{T}} \phi(R_{n}k_n)v_n}{\Sigma_{n=1}^{N} \phi(R_{m}q_m)^{\mathsf{T}}\phi(R_{n}k_n)}
\end{equation} 
where $R_m$ represents a position-dependent rotation at position $m$, and $\phi(x) = \text{elu}(x) + 1$ defines the activation function used. This linear attention mechanism reduces the complexity to $O(N)$, significantly improving the efficiency of chemical language representation learning, coming with minimal performance degradation. 

\subsection{Chemical Language Representation Learning via MLM}
Typical work in chemical language representation learning \citep{Chithrananda-Neuripsworkshop2020,Ross-NMI2022} utilizes a self-supervised task known as MLM. This objective involves training a model to predict original sequences from sequences in which some tokens are randomly masked. Specifically, given a sequence $\textbf{X}=\{x_1,x_2,x_3,...,x_n\}$, we corrupt $\textbf{X}$ into $\tilde{\textbf{X}}$ by masking 15\% of its tokens. We then train a model, denoted as $C$ with parameters $\theta_{C}$, to reconstruct $\textbf{X}$. The loss of each example is formulated as follows:
\begin{equation}
    \mathcal{L}_{C} = - \sum_{i \in \mathcal{M}} \log p(x_i|\tilde{\textbf{X}};\theta_{C}),
\end{equation}
where $\mathcal{M}$ represents the set of masked token positions. In typical chemical language representation learning methods, each masked token is substituted with a special mask token in 80\% of cases, a random token in 10\% of cases, and the original token in the remaining 10\% cases, following practices in BERT \cite{Devlin-naacl2019}.

\section{MolTRES: Molecular Transformer with Enhanced Self-supervised Learning}
In this section, we detail our framework, MolTRES. We propose a novel pre-training task, called DynaMol, which incorporates generator-discriminator training into chemical language representation learning with substructure masking. In addition, we integrate molecular representations that have been trained on scientific literature.

\subsection{DynaMol: Dynamic Molecule Modeling with Generator-Discriminator Training}
To increase the difficulty of chemical language representation learning, we propose a dynamic molecule modeling scheme based on generator-discriminator training, inspired by replaced token detection proposed in \citet{Clark-ICLR2020}. The proposed scheme involves training two models, namely a generator and a discriminator. The generator is trained to predict original sequences given masked sequences similar to MLM, while the discriminator is trained to identify tokens that have been replaced by the generator. Since the generator transforms masked sequences to more closely resemble original distributions, this training scheme results in less discrepancy between the inputs from pre-training and downstream tasks, and allows for flexible adjustments of the masking ratio \cite{he2022debertav3}. Moreover, as the generator is being trained, it naturally provides increasingly challenging examples to the discriminator. This scheme is expected to alleviate the issues of early convergence and over-fitting commonly observed in existing methods of chemical language representation learning.

Specifically, similar to MLM, the generator $G$ with parameters $\theta_{G}$ is trained to reconstruct the sequence $\textbf{X}$. The loss of $G$ for each example is formulated as follows:
\begin{equation}
    \mathcal{L}_{G} = - \sum_{i \in \mathcal{M}} \log p(x_i|\tilde{\textbf{X}};\theta_{G}).
\end{equation}

Then, the input sequence for the discriminator is constructed by replacing the masked tokens in $\tilde{\textbf{X}}$ with new tokens, sampled from the generator's probability distribution $p_{G}$, as follows:
\begin{equation}
    \tilde{\textbf{X}}_{D} =  
\begin{cases}
    \tilde{x}_{i} \backsim p(x_i|\tilde{\textbf{X}};\theta_{G}), & \text{if } i \in \mathcal{M} \\
    x_i,              & \text{otherwise.}
\end{cases}
\end{equation}

The discriminator is trained to distinguish whether each token in the generated input sequence $\tilde{\textbf{X}}_{D}$ is original or has been replaced. The loss for the discriminator is formulated as follows:
\begin{equation}
    \mathcal{L}_{D} = -\sum_{i=1}^{n} \log p (z_i|\tilde{\textbf{X}}_{D};\theta_{D}),
\end{equation}
where $z_i$ is a binary label that indicates whether the $i$-th input token is original or has been replaced. Finally, the generator $G$ and discriminator $D$ are jointly optimized with multiple objectives, expressed as $\mathcal{L} = \mathcal{L}_{G} + \lambda \mathcal{L}_{D}$, where $\lambda$ is a pre-defined balancing parameter for the discriminator loss. In this work, $\lambda$ is set to 10.


In addition, we carefully design three rules to mask SMILES at multiple substructure-level granularities, thereby preventing models from predicting the correct answer by exploiting superficial patterns in the SMILES grammar. (1) We mask all special tokens that represent structural information, such as numbers for cycles. (2) We then mask spans of SMILES that composes certain substructures, such as substituents, bridges, or groups of sequential atoms, until the ratio of masked tokens does not exceed the pre-defined target masking ratio. Note that these substructure can be easily identified by segmenting SMILES strings based on brackets. (3) Finally, we mask random atomic SMILES tokens to achieve the target masking ratio. We follow a typical masking strategy that, among the masked tokens, 80\% are replaced with mask tokens, 10\% are replaced with random tokens, and the rest 10\% remain unchanged. Notably, we use 65\% of the target masking ratio for pre-training.


\begin{table*}[t!]
\small
\centering
\setlength{\tabcolsep}{3pt}
\begin{tabular}{lcccccccccc}
\toprule
Methods         & BBBP $\uparrow$   & Tox21 $\uparrow$    & ToxCast $\uparrow$    & ClinTox $\uparrow$    & MUV $\uparrow$    & HIV $\uparrow$    & BACE $\uparrow$    & SIDER $\uparrow$  & Avg. $\uparrow$ \\ \midrule
\textbf{3D Conformation} \\
GeomGCL  \citep{3dgraph}                & -     & \underline{85.0}  & -     & 91.9  & -     & -     & -     & 64.8 & - \\
GEM   \citep{gem}                   & 72.4  & 78.1  & -     & 90.1  & -     & 80.6  & 85.6  & 67.2  & - \\
3D InfoMax \citep{3dinfomax} & 68.3 & 76.1 & 64.8 & 79.9 & 74.4 & 75.9 & 79.7 & 60.6 & 72.5\\
GraphMVP \citep{3dgraph} & 69.4  & 76.2  & 64.5 & 86.5 & 76.2 & 76.2  & 79.8  & 60.5 & 73.7 \\
MoleculeSDE \citep{moleculesde} & 71.8 & 76.8 &  65.0 & 87.0 & 80.9 & 78.8 & 79.5 & 75.1 \\ 
Uni-Mol \citep{zhou2022uni} & 71.5 & 78.9 & 69.1 & 84.1 & 72.6 & 78.6 & 83.2 & 57.7 & 74.5\\
MoleBlend \citep{moleblend} & 73.0 & 77.8 & 66.1     & 87.6 & 77.2    & 79.0 & 83.7 & 64.9 & 76.2 \\
Mol-AE \citep{yang2024mol} & 72.0 & 80.0 & \underline{69.6} & 87.8 & 81.6 & 80.6 & 84.1 & 67.0 & 77.8\\
UniCorn \citep{feng2024unicorn} & 74.2 & 79.3 & 69.4 & 92.1 & \underline{82.6} & 79.8 & 85.8 & 64.0 & 78.4\\\midrule
\textbf{2D Graph} \\
DimeNet \citep{dimenet}                 & -     & 78.0  & -     & 76.0  & -     & -     & -     & 61.5 & - \\ 
AttrMask \citep{Hu-iclr2020} & 65.0  & 74.8  & 62.9 & 87.7  & 73.4 & 76.8  & 79.7  & 61.2 & 72.7 \\
GROVER \citep{grover} & 70.0 & 74.3 & 65.4 & 81.2 & 67.3 & 62.5 & 82.6 & 64.8 & 71.0\\
BGRL \citep{thakoor2021large} & 72.7 & 75.8 & 65.1 & 77.6 & 76.7 & 77.1 & 74.7 & 60.4 & 72.5 \\
MolCLR  \citep{Wang-NatMachIntell2022}                 & 66.6  & 73.0 & 62.9 & 86.1  & 72.5 & 76.2  & 71.5  & 57.5 & 70.8 \\
GraphMAE \citep{graphmae} & 72.0 & 75.5 & 64.1 & 82.3 & 76.3 & 77.2 & 83.1 & 60.3 & 73.9 \\
Mole-BERT \citep{liu2023rethinking} & 71.9 & 76.8 & 64.3 & 78.9 & 78.6 & 78.2 & 80.8 & 62.8 & 74.0 \\
SimSGT \citep{molebert} & 72.2 & 76.8 & 65.9 & 85.7 & 81.5 & 78.0 & 84.3 & 61.7 & 75.8 \\
MolCA + 2D \citep{liu2023molca} & 70.0  & 77.2  & 64.5 & 89.5  & - & -  & 79.8 & 63.0 & - \\ \midrule
\textbf{1D SMILES/SELFIES} \\
MoLFormer-XL \citep{Ross-NMI2022}  & 93.7  & 84.7  & 65.6 & \underline{94.8}  & 80.6  & \underline{82.2}  & \underline{88.2}  &  66.9 & \underline{82.1} \\
SELFormer \citep{selformer}        & 90.2  & 65.3  & -     & -  & - & 68.1  & 83.2  &  \textbf{74.5} & - \\
MolCA \citep{liu2023molca} & 70.8  & 76.0  & 56.2 & 89.0  & - & -  & 79.3 & 61.2  & -  \\
MolTRES-small (ours)      & \underline{95.0} & 83.4 & 64.8 & 94.0 & 80.0 & 81.7 & 87.7 & 68.3 & 81.9\\
MolTRES (ours)      & \textbf{96.1} & \textbf{85.3} & \textbf{70.1}     & \textbf{96.7} &   \textbf{84.9}    & \textbf{84.2} & \textbf{91.7} & \underline{69.8} & \textbf{84.8}\\ \bottomrule
\end{tabular}
\caption{Evaluation results on MoleculeNet classification tasks. We report ROC-AUC scores (higher is better) under scaffold splitting. The best and second-best results are in \textbf{bold} and \underline{underlined}.}
\label{classification_results}
\end{table*}

\subsection{Knowledge Transfer from Scientific Literature using mat2vec} 
While modeling SMILES helps models understand molecular structure and connectivity, SMILES itself lacks explicit information about molecular properties. Scientific literature, which is similarly represented in a textual form, provides a more flexible and rich source of external information. It comprehensively involves information about molecular properties derived from wet laboratory experiments and computational methods. Therefore, we enrich the representations of SMILES Transformers by integrating information from scientific literature. 

Despite the many possible design choices available, we opt to leverage mat2vec \citep{Tshitoyan-Nature2019}, a straightforward embedding model trained on extensive scientific literature, for integration into Transformer's embedding vectors. We prioritize the efficiency in terms of memory footprints and computations in our integration procedures, essential for dealing with large-scale pre-training. Given an input sequence $\textbf{X}=\{x_1,...,x_n\}$, we obtain embedding vectors for every token from the Transformer's embedding layer, denoted as $\textbf{E}^t=\{e^t_1,...,e^t_n\}$. Using a mapping function $I(\cdot)$, we assign each token to corresponding mat2vec embedding vectors, denoted as $\textbf{E}^m=\{e^m_1,...,e^m_n\} \text{ s.t. } e^m_k = \sum_{z \in I(x_k)} \text{mat2vec}(z)$. We then combine $\textbf{E}^t$ and $\textbf{E}^m$ using a linear projection layer $F_{1}(\cdot)$. The set of embedding vectors for the generator $V_G$ is generated as follows:
\begin{equation}
\begin{split}
    \textbf{V}_{G} =& \{ F_{1}(e^t_1 \circ e^m_1), ..., F_{1}(e^t_n \circ e^m_n) \},
\end{split}
\end{equation}
where $\circ$ denotes the concatenation operation. In a similar manner, the set of embedding vectors for the discriminator $V_D$ is generated from the tokens reconstructed by the generator as follows:
\begin{equation}
\begin{split}
    \textbf{V} =& \{F_{1}(\tilde{e}^t_1 \circ \tilde{e}^m_1), ..., F_{1}(\tilde{e}^t_n \circ \tilde{e}^m_n) \}\\
    \textbf{V}_{D} =& \{ F_{2}(\sigma(v_1)), ..., F_{2}(\sigma(v_n)) \}\\
    &\text{s.t. } v_1,...,v_n \in V,
\end{split}
\end{equation}
where $v_1,...,v_n \in V$ and $\sigma(\cdot)$ is an activation function, which is the gelu function in this work.

For the integration, we manually design a mapping function $I(\cdot)$ using human prior knowledge to address the vocabulary mismatch between SMILES tokens and mat2vec words. We utilize a thesaurus carefully constructed by domain experts, chosen for its superior computational efficiency and stability compared to learning-based approaches. For example, the thesaurus maps ``[cH+]'' in the Transformer's vocabulary to ``methylidyne'', ``ion'', and ``cation'' in the mat2vec vocabulary. Based on this thesaurus, we pre-calculate embedding vectors for 2,696 tokens in the Transformer vocabulary before pre-training. To prevent catastrophic forgetting of mat2vec knowledge, we freeze these pre-calculated embedding vectors during pre-training. During fine-tuning, these embedding vectors are trainable to adapt the knowledge for each downstream task.

\begin{table*}[t!]
\small
\centering
\begin{tabular}{lccccc}
\toprule
Methods        & ESOL $\downarrow$ & FreeSolv $\downarrow$ & Lipophilicity $\downarrow$ & Avg. $\downarrow$ \\ \midrule
\textbf{3D Conformation} \\
3D InfoMax \citep{3dinfomax} & 0.894 & 2.337 & 0.695 & 1.309 \\
GraphMVP \citep{3dgraph} & 1.029  & -  & 0.681 & - \\
Uni-Mol \citep{zhou2022uni} & 0.844 & 1.879 & 0.610 & 1.111 \\
MoleBlend \citep{moleblend} & 0.831 & 1.910 & 0.638 & 1.113 \\
Mol-AE \citep{yang2024mol} & 0.830 & 1.448 & 0.607 & 0.962\\
UniCorn \citep{feng2024unicorn} & 0.817 & 1.555 & 0.591 & 0.988 \\\midrule
\textbf{2D Graph} \\
AttrMask \citep{Hu-iclr2020} & 1.112  & - & 0.730 & - \\
GROVER \citep{grover} & 0.831 & 1.544 & 0.560 & 0.978 \\
MolCLR  \citep{Wang-NatMachIntell2022} & 1.110  & 2.200  & 0.650 & 1.320\\
SimSGT \citep{liu2023rethinking} & 0.917 & - & 0.695 & - \\
\midrule
\textbf{1D SMILES/SELFIES} \\
MoLFormer-Base \citep{Ross-NMI2022} & 0.280   & 0.260  & 0.649 & 0.396 \\
MoLFormer-XL   \citep{Ross-NMI2022} & \underline{0.279}        & \underline{0.231}  & \underline{0.530}  & \underline{0.347}  \\
SELFormer  \citep{selformer} & 0.682        & 2.797        & 0.735  & 1.405  \\
MolTRES-small (ours) & 0.280 & 0.250 & 0.594 & 0.375 \\
MolTRES (ours) & \textbf{0.274} & \textbf{0.229} & \textbf{0.504} & \textbf{0.336} \\ \bottomrule
\end{tabular}
\caption{Evaluation results on MoleculeNet regression tasks. We report RMSE scores (lower is better) under scaffold splitting. The best and second-best results are in \textbf{bold} and \underline{underlined}.}
\label{regression_results}
\end{table*}

\section{Experiment}

\subsection{Experimental Setup}

\paragraph{Pre-training.} We collect 118 million molecules from PubChem\footnote{https://pubchem.ncbi.nlm.nih.gov/} and 1.9 billion molecules from ZINC\footnote{https://zinc.docking.org/}. We pre-train two MolTRES models, a base model (MolTRES) and a smaller model (MolTRES-small). Our model architectures are detailed in Appendix \ref{app:exp}. We train our models for 200,000 steps with a batch size of 25,600 and use the final models in evaluation.

\paragraph{Evaluation} We evaluate our models and baselines on eight classification tasks and four regression tasks from the MoleculeNet benchmark \citep{wu2018moleculenet}. We report Receiver Operating Characteristic-Area Under the Curve (ROC-AUC) scores for the classification tasks, Mean Absolute Error (MAE) scores for QM9, and Root Mean Square Error (RMSE) scores for the remaining regression tasks. We report the test score from the model that achieves the best validation score.

\paragraph{Baselines.} We compare our models with diverse state-of-the-art baselines categorized as follows:
\begin{itemize}
    \item \textbf{3D Conformation:} This category includes methods that utilize 3D conformation from the geometry information of molecules and may incorporate other modalities.
    \item \textbf{2D Graph:} This category includes methods that utilize 2D graph information, such as atoms and bonds, and may also combine 1D SMILES. 
    \item \textbf{1D SMILES/SELFIES:} This category includes methods that utilize SMILES or SELFIES sequences of molecules.
\end{itemize}

\begin{table*}[t!]
\centering
\resizebox{\textwidth}{!}{%
\setlength{\tabcolsep}{3pt}
\begin{tabular}{lcccccccccccccc}
\toprule
Methods & $\mu\downarrow$ & $\alpha\downarrow$ & $\varepsilon_{homo}\downarrow$ & $\varepsilon_{lumo}\downarrow$ & $\Delta\varepsilon\downarrow$ & $\langle R^2\rangle\downarrow$ & $ZPVE\downarrow$ & $U_0\downarrow$ & $U_{298}\downarrow$ & $H_{298}\downarrow$  & $G_{298}\downarrow$ & $C_v\downarrow$ & Avg.$\downarrow$ \\
& (D) & ($a_0^3$) & (eV) & (eV) & (eV) & ($a_0^2$) & (eV) & (eV) & (eV) & (eV)  & (eV) & ($\frac{\text{cal}}{\text{mol} \cdot \text{K}}$) & \\
\midrule
\rowcolor{Gray}\textbf{3D Conformation (GT)} &&&&&&&&&&&&& \\
\rowcolor{Gray}3D InfoMax \cite{3dinfomax} & 0.028 & 0.057 & 0.259 & 0.216 & 0.421 & \underline{0.141} & 0.002 & 0.013 & 0.014 & 0.014 & 0.014 & 0.030 & 0.101 \\
\rowcolor{Gray}GraphMVP \citep{3dgraph} & 0.030 & 0.056 & 0.258 & 0.216 & 0.420 & \textbf{0.136} & 0.002 & 0.013 & 0.013 & 0.013 & 0.013 & 0.029 & 0.100 \\
\rowcolor{Gray}MoleculeSDE \cite{moleculesde} & \underline{0.026} & \underline{0.054} & 0.257 & 0.214 & 0.418 & 0.151 & 0.002 & 0.012 & 0.013 & 0.012 & 0.013 & \underline{0.028} & \underline{0.100} \\
\rowcolor{Gray}MoleBlend \cite{moleblend} & 0.037 & 0.060 & \underline{0.215} & \underline{0.192} & \underline{0.348} & 0.417 & \underline{0.002} & \underline{0.012} & \underline{0.012} & \underline{0.012} & \underline{0.012} & 0.031 & 0.113 \\
\rowcolor{Gray}UniCorn \cite{feng2024unicorn} & \textbf{0.009} & \textbf{0.036} & \textbf{0.130} & \textbf{0.120} & \textbf{0.249} & 0.326 & \textbf{0.001} & \textbf{0.004} & \textbf{0.004} & \textbf{0.004} & \textbf{0.005} & \textbf{0.019} & \textbf{0.076} \\
\midrule\midrule
\textbf{3D Conformation (RDKit)} \\
SchNet \cite{schutt2017schnet} & 0.447 & 0.276 & 0.082 & 0.079 & 0.115 & 21.58 & 0.005 & 0.072 & 0.072 & 0.072 & 0.069 & 0.111 & 1.915 \\
3D InfoMax \cite{3dinfomax} & 0.351 & 0.313 & 0.073 & 0.071 & 0.102 & 19.16 & 0.013 & 0.133 & 0.134 & 0.187 & 0.211 & 0.165 & 1.743 \\
MoleculeSDE \cite{moleculesde} & 0.423 & \underline{0.255} & 0.080 & 0.076 & 0.109 & 20.43 & \textbf{0.004} & \textbf{0.054} & \textbf{0.055} & \textbf{0.055} & \textbf{0.052} & \underline{0.098} & 1.808 \\
\midrule
\textbf{2D Graph} \\
1-GNN \cite{morris2019weisfeiler} & 0.493 & 0.780 & 0.087 & 0.097 & 0.133 & 34.10 & 0.034 & 63.13 & 56.60 & 60.68 & 52.79 & 0.270 & 22.43 \\
1-2-3-GNN \cite{morris2019weisfeiler} & 0.476 & 0.270 & 0.092 & 0.096 & 0.131 & 22.90 & \underline{0.005} & 1.162 & 3.020 & 1.140 & 1.276 & \textbf{0.094} & 2.012 \\
\midrule
\textbf{1D SMILES/SELFIES} \\
MoLFormer-XL \cite{Ross-NMI2022} & 0.362 & 0.333 & 0.079 & 0.073 & 0.103 & 17.06 & 0.008 & 0.192 & 0.245 & 0.206 & 0.244 & 0.145 & 1.588  \\
MolTRES-small (ours) & \underline{0.326} & 0.295 & \underline{0.066} & \underline{0.067} & \underline{0.085} & \underline{16.32} & 0.009 & 0.133 & 0.185 & 0.155 & 0.164 & 0.137 & \underline{1.495} \\
MolTRES (ours) & \textbf{0.315} & \textbf{0.237} & \textbf{0.054} & \textbf{0.057} & \textbf{0.077}  & \textbf{14.60} & 0.007 & \underline{0.061} & \underline{0.071} & \underline{0.068} & \underline{0.057} & 0.121  & \textbf{1.310}\\ \bottomrule
\end{tabular}%
}
\caption{Evaluation results on QM9 tasks. We report MAE scores (lower is better) following the data splitting used in \citet{moleculesde}. The best and second-best results are in \textbf{bold} and \underline{underlined}. It is important to note that the ``3D Conformation (GT)'' results utilize ground-truth geometry information, which incurs non-trivial costs to obtain. For a fair comparison, we also evaluate the performance of 3D models using the geometry information approximated by RDKit, denoted as ``3D Conformation (RDKit)'', considering scenarios where ground-truth geometry is unavailable.}
\label{qm9}
\end{table*}

\subsection{Main Results}
We first compare MolTRES with state-of-the-art molecular property prediction methods on MoleculeNet classification tasks. As shown in Table \ref{classification_results}, MolTRES surpasses the best baseline, MoLFormer-XL, by an average of 2.7\%. In addition, MolTRES-small also shows a competitive performance compared to the baselines. Notably, MolTRES significantly outperforms baseline methods using 3D conformation and 2D graph. This confirms the strength of pre-training with billion-scale SMILES sequences, compared to pre-training with hundreds of millions of conformation or graph examples. MolTRES exhibits state-of-the-art performance on 7 of the 8 tasks. Although MolTRES achieves the second-best results after SELFormer on the SIDER task, it outperforms SELFormer by up to 20\% on the others, affirming the superiority of MolTRES.

Moreover, as shown in Table \ref{regression_results}, MolTRES consistently stands out in three MoleculeNet regression tasks, surpassing the state-of-the-art method MoLFormer-XL by an average of 3.3\%. Moreover, MolTRES-small achieves better performance than MoLFormer-Base, which contains a commensurate number of parameters, by an average of 5.6\%. The superior performance of SMILES-based methods is still observed, as they achieve significantly smaller errors compared to other baseline methods. This performance gap further verifies the efficacy of large-scale pre-training on SMILES. 


We further compare MolTRES with the baselines on QM9, as shown in Table \ref{qm9}. Since quantum properties are strongly correlated with geometry information, baselines using ground-truth geometry information (3D Conformation (GT)) show the best results among baselines. However, obtaining this geometry information involves non-trivial costs and may not be available in many real-world scenarios. In these contexts, our MolTRES models provide the most accurate approximation by only using SMILES, compared to baselines that estimate geometry information from RDKit or those without any geometry information, demonstrating its efficacy and applicability.

\subsection{Analysis}
To better understand the performance improvements from MolTRES, we conduct a series of analysis on four MoleculeNet classification tasks: BBBP, ClinTox, BACE, and SIDER.

\paragraph{Ablation Study.} To assess the distinct contributions of MolTRES's components to its enhanced performance, we conduct ablation studies using variants of MolTRES as detailed in Table \ref{tab:ablation}. The results demonstrate that both the DynaMol and mat2vec integration contribute to performance improvements. Moreover, when used jointly, they offer complementary advantages over employing either method in isolation. This result underscores MolTRES's effectiveness in addressing the issues in existing chemical language representation learning, leading to notable performance improvements.

\begin{figure*}[t]
\centering
\includegraphics[width = 0.49\linewidth]{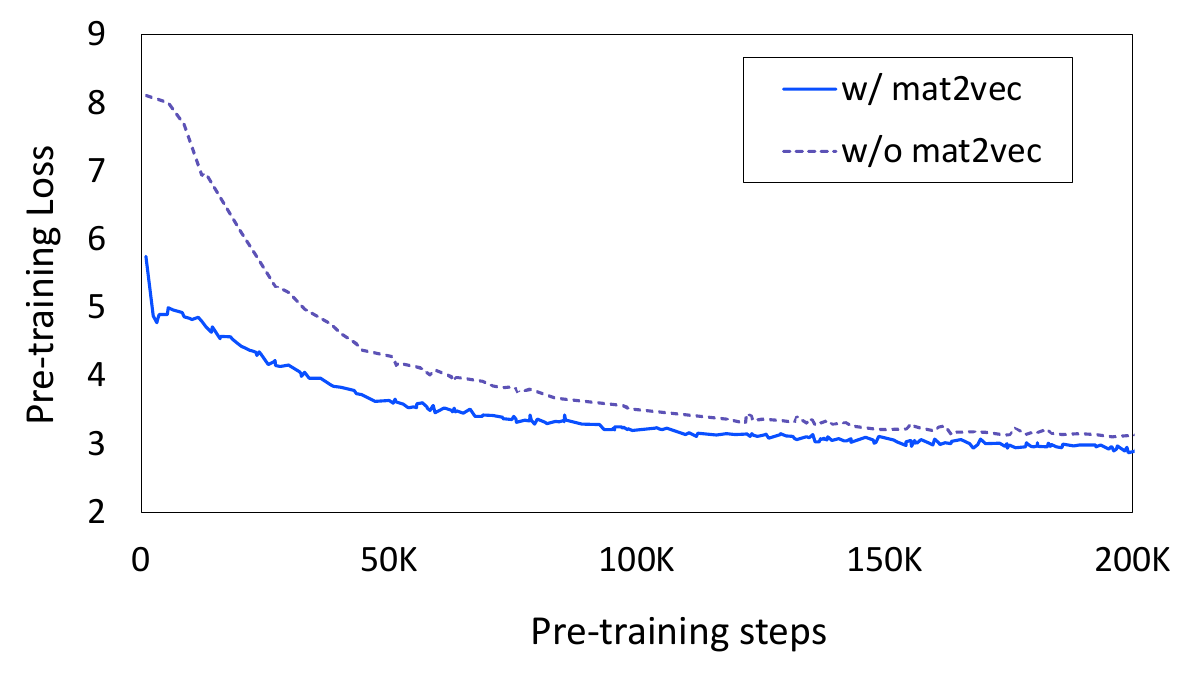}
\includegraphics[width = 0.49\linewidth]{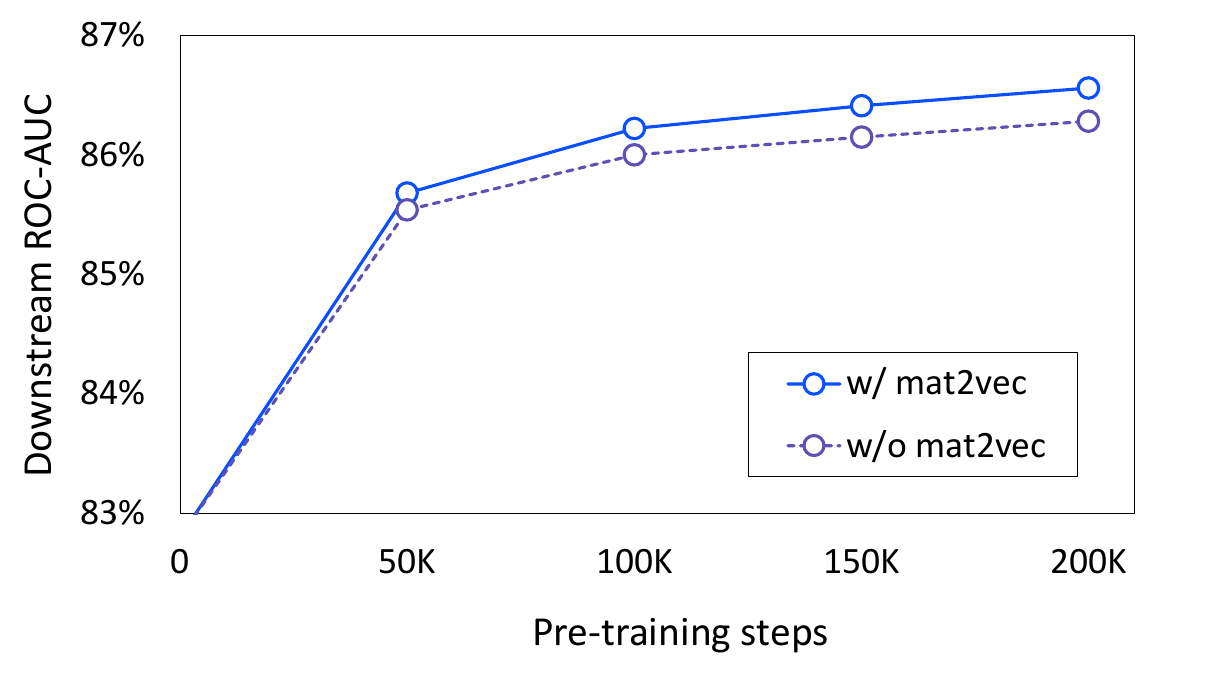}
\caption{Training curves of MolTRES with mat2vec embeddings (the solid line) and without mat2vec embeddings (the dashed line). The left shows the pre-training loss curves, while the right shows the average ROC-AUC scores.}
\label{fig:m2v}
\end{figure*}

\paragraph{Effect of mat2vec embedding.} We analyze the effect of the mat2vec embeddings on the pre-training of MolTRES. As described in Figure \ref{fig:m2v}, mat2vec enables faster convergence, attributed to the rich features provided by mat2vec that are beneficial for structure modeling. Additionally, when fully trained, MolTRES with mat2vec achieves lower training losses and enhanced performance in MoleculeNet classification tasks. This validates the effectiveness of  integrating mat2vec embeddings.


\begin{table}[t]
\centering
\begin{small}
\begin{tabular}{cc|ccc}
\toprule
DynaMol & mat2vec & ROC-AUC $\uparrow$ \\ \midrule
\checkmark & \checkmark & \textbf{87.99} \\
\checkmark & - & 87.67\\
- & \checkmark & 84.82 \\
- & - & 84.05 \\ \bottomrule
\end{tabular}
\caption{Performance on MoleculeNet classification tasks with variants of MolTRES.}
\label{tab:ablation}
\end{small}
\end{table}

\section{Related Work}
In recent years, representation learning has prevailed in numerous applications in natural language processing \citep{Devlin-naacl2019, Liu-corr2019} and computer vision \cite{dosovitskiy2020image,bao2021beit}. This trend has triggered many studies in chemical representation learning. The approaches in this field can be classified into three categories based on molecular descriptors used for pre-training: chemical language representation learning, chemical graph representation learning, and multi-modal chemical representation learning.

\paragraph{Chemical language representation learning.} 
Chemical language representation learning has adopted pre-training on molecular descriptors represented as strings, such as SMILES and SELFIES. It typically leverages Transformers \citep{Vaswani-NIPS2017} to learn molecular descriptors inspired by the recent success of large-scale representation learning in natural language processing. \citet{Wang-BCB2019,Chithrananda-Neuripsworkshop2020,Ross-NMI2022} have trained Transformer models on large-scale SMILES sequences. \citet{selformer} have utilized SELFIES sequences to achieve a better representation space. However, the training strategies for these methods follow the practice of MLM-style training in natural language processing. Since chemical language differs from natural language, current applications of MLM encounter various issues in pre-training. In this work, we propose MolTRES to address these issues and consequently improve molecular property prediction.


\paragraph{Chemical graph representation learning.} Researchers in chemical graph representation learning argue that molecules can naturally be represented in 2D or 3D graph structures. Thus, they typically leverage graph neural networks (GNNs) or Transformers adapted to graphs. \citet{Hu-iclr2020} have introduced a self-supervised task for molecular graphs, called AttrMask. \citet{Morris-AAAI2019} have introduced higher-order GNNs for distinguishing non-isomorphic graphs. \citet{You-Neurips2020} have extended contrastive learning to unstructured graph data. \citet{Wang-NatMachIntell2022} have proposed a unified GNN pre-training framework that integrates contrastive learning and sub-graph masking. Recent work has focused on modeling 3D graphs, as they provide more vital information for predicting molecular properties compared to 2D graphs. \cite{yang2024mol,zhou2022uni} have proposed denoising auto-encoders for directly modeling 3D graphs. However, due to the limited scale of 3D molecular data and its resource-intensive modeling, the applicability of 3D approaches is limited.

\paragraph{Multi-modal chemical representation learning.} Recently, several studies have proposed learning chemical representations in a multi-modal manner, typically leveraging both 2D topology and 3D geometry of molecules. \citet{3dgraph,3dinfomax,moleculesde} have introduced a contrastive learning framework that uses 2D graphs and their corresponding 3D conformations as positive views, treating those from different molecules as negative views. \cite{luo2022one} have proposed encoding both 2D and 3D inputs within a single GNN model. Another research direction has involved using both chemical and natural languages \cite{edwards2022translation,liu2023molca} to enrich molecular representations and facilitate molecule generation using natural language. We plan to further explore the multi-modal and generation capabilities of MolTRES based on its versatile Transformer architecture.

\section{Conclusion}
In this work, we have proposed a novel chemical language representation learning framework, MolTRES, to address the limited scalability and generalizability of existing methods for pre-training SMILES transformers. We have presented two methods, dynamic molecule modeling with generator-discriminator training, called DynaMol, and knowledge transfer from scientific literature based on mat2vec. Our experimental results validate the superiority of our framework over existing chemical models across a wide range of molecular property prediction tasks. 

\section*{Limitations}
While we have demonstrated that MolTRES effectively improves molecular property prediction by addressing issues in existing chemical language representation learning methods, some limitations open promising avenues for future research. First, several components in MolTRES, such as its masking strategy or knowledge transfer method, were chosen empirically in terms of efficiency, and therefore may have room for performance improvements through theoretical or learning-based approaches. Second, we evaluated a few architectural settings of MolTRES corresponding to those of MoLFormer-XL for comparison. Future evaluations could explore more diverse settings of MolTRES to accommodate various scenarios, including resource-limited or scalable environments. Finally, a popular application of SMILES Transformers is in molecule generation. We plan to investigate the extension of MolTRES on the pre-training of generative Transformers for this purpose.



\bibliography{acl_latex}

\appendix

\section{Appendix}
\label{sec:appendix}

\begin{table*}[ht!]
\centering
\begin{small}
\begin{tabular}{clcc}
\hline
 & Descriptions & \# tasks & \# samples \\ \hline
BBBP & Blood brain barrier penetration dataset & 1 & 2,039 \\
Tox21 & Toxicity measurements on 12 different targets & 12 & 7,831 \\
ToxCast & Toxicology data for a large library of compounds & 617 & 8,577 \\
Clintox & Clinical trial toxicity of drugs & 2 & 1,478 \\
MUV & Maximum unbiased validation group from PubChem BioAssay & 17 & 93,087 \\
HIV & Ability of small molecules to inhibit HIV replication & 1 & 41,127 \\
BACE & Binding results for a set of inhibitors for  $\beta-$ secretase 1 & 1 & 1,513 \\
SIDER & Drug side effect on different organ clases & 27 & 1,427 \\ \hline
\end{tabular}
\end{small}
\caption{Classification tasks from MoleculeNet.}
\label{classification_statistics}
\end{table*}

\begin{table*}[ht!]
\centering
\begin{small}
\begin{tabular}{clcc}
\hline
 & Descriptions & \# tasks & \# samples \\ \hline
QM9 & 12 quantum mechanical calculations of organic molecules & 12 & 133,885 \\
ESOL & Water solubility dataset & 1 & 1,128 \\
FreeSolv & Hydration free energy of small molecules in water & 1 & 642 \\
Lipophilicity & Octanol/water distribution coefficient of molecules & 1 & 4,200 \\ \hline
\end{tabular}
\end{small}
\caption{Regression benchmarks from MoleculeNet.}
\label{regression_statistics}
\end{table*}

\begin{table*}[t!]
\small
\centering
\begin{tabular}{ccccccc}
\toprule
& \multicolumn{2}{c}{Generator} & \multicolumn{2}{c}{Discriminator} & ROC-AUC $\uparrow$ & MAE $\downarrow$ \\ 
& \# layers & Hidden size & \# layers & Hidden size & (CLS) & (REG) \\ \midrule
\multirow{2}{*}{(A)} & 3 &  & 6 &  &  81.9 & 0.375 \\
&  & 512 & & 512 &  82.2 & 0.371\\\midrule
\multirow{2}{*}{(B)} & 12 & 384 & &  &  83.6 & 0.341 \\
& 12 & 512 &  &  &  84.7 & \textbf{0.336}\\\midrule
\multirow{3}{*}{(C)} & 4 & &  &  &  83.3 & 0.343 \\
& 8 &  &  &  &  84.5 & \textbf{0.336} \\
& 12 &  &  &  &  84.0 & 0.337\\\midrule
(D) & 6 & 768 & 12 & 768 &  \textbf{84.8} & \textbf{0.336} \\ \bottomrule
\end{tabular}
\caption{Variations on the MolTRES architectures. Unlisted values are identical to those of the standard setting of MolTRES in (D). Following the experimental settings described in Section 4.1, ROC-AUC scores are measured on eight MoleculeNet classification tasks and MAE scores are measured on three MoleculeNet regression tasks.}
\label{ablations}
\end{table*}

\subsection{Detailed Experimental Settings}\label{app:exp}

\paragraph{Pre-training.} For pre-processing, we extract the canonicalized format of SMILES for every molecule using RDKit. We construct the vocabulary with 2,691 unique tokens plus five special tokens (``<bos>'', ``<eos>'', ``<pad>'', ``<mask>'', and ``<unk>'') after tokenizing all the extracted SMILES sequences. For tokenization, we use the maximum sequence length of 512. The weights of our models are initialized over the normal distribution with a standard deviation of 0.02.  Pre-training is performed using an AdamW optimizer ($\beta_1=0.9$, $\beta_2=0.95$), where the maximum learning rate and weight decay are set to 3e-4 and 0.01, respectively. We use the cosine annealing for learning rate scheduling with 1,000 warmup steps. The pre-training time of MolTRES is approximately 15 days using 4 NVIDIA RTX A6000 GPUs. 

\paragraph{Evaluation} The statistics of evaluation benchmarks are shown in Table \ref{classification_statistics} and \ref{regression_statistics}. We use the scaffold splitting (80\% / 10\% / 10\% for train / validation / test) for all the tasks except for QM9, in which the random split (80\% / 10\% / 10\% for train / validation / test) with thermochemical energy pre-calculation is used following \citet{moleculesde}. For evaluation of our models, we extract the output representations from model's final transformer block corresponding to the first input token (``<bos>'') as the molecule representations. We use a 2-layer MLP with the same hidden size and gelu activation for prediction, whose weights are initialized over the normal distribution with a standard deviation of 0.02. We use the augmentation of random SMILES reconstruction for all the tasks. We fine-tune the models for 500 epochs using an AdamW optimizer ($\beta_1=0.9$, $\beta_2=0.99$) with a weight decay of 0.01. For each task, we empirically choose the batch size $\in \{16,32,64,128\}$ and learning rate $\in \{\text{2e-5}, \text{3e-5}, \text{5e-5}, \text{1e-4}\}$. We report the average scores after five runs.

\paragraph{Model Architecture.} The model architecture of the generator and discriminator is a Transformer with linear attention and rotary position embeddings. The discriminator of MolTRES has 12 layers, 768 hidden dimensions, and 12 attention heads. The discriminator of MolTRES-small has 6 layers, 768 hidden dimensions, and 12 attention heads. The generators have half the number of layers in their corresponding discriminator, while the other settings are consistent. It is noteworthy that the generator is only used for pre-training, and the discriminator is fine-tuned and evaluated in all the downstream tasks. The generator and discriminator share their embeddings, which is known to be beneficial in accelerating the pre-training \cite{Clark-ICLR2020}.

\begin{figure}[t]
\centering
\includegraphics[width = \linewidth]{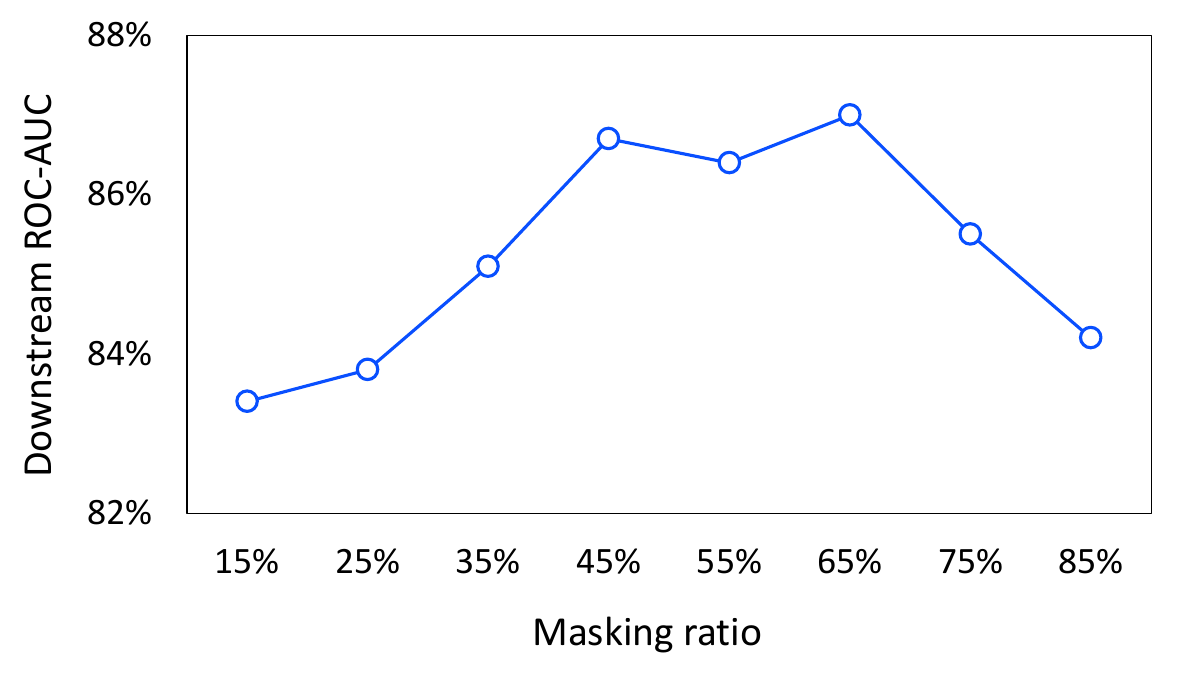}
\caption{Comparison of MolTRES for different masking ratios on MoleculeNet classification tasks.}
\end{figure}
\begin{figure}[t]
\centering
\includegraphics[width = \linewidth]{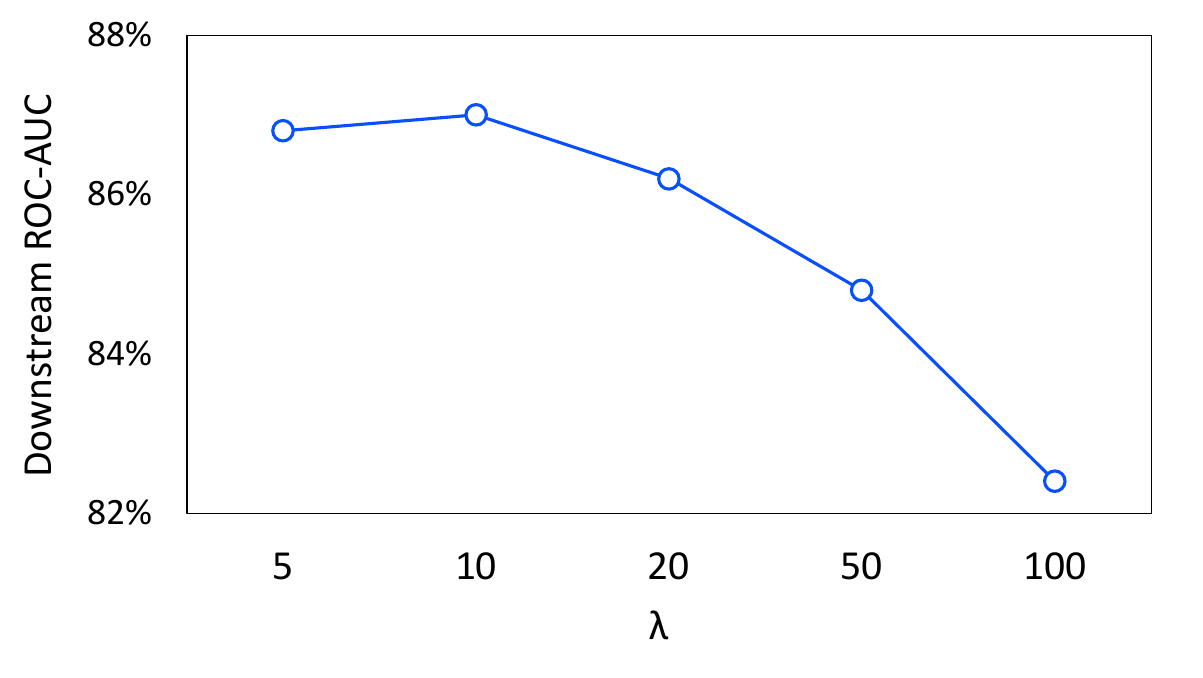}
\caption{Comparison of MolTRES for different $\lambda$ on MoleculeNet classification tasks.}
\end{figure}

\subsection{Additional Experimental Results}

\paragraph{Pre-training hyper-parameter analysis.} We study the effect of pre-training hyper-parameters as shown in Figures 4 and 5. We report ROC-AUC scores on four MoleculeNet classification tasks (BBBP, ClinTox, BACE, and SIDER). First, in Figure 4, we find that the optimal masking ratio for MolTRES is 65\%. When the masking ratio is smaller than 65\%, we observe that the generator easily fills masked tokens, resulting in significantly biased labels towards original. In contrast, when the masking ratio is larger than 65\%, we observe that there is few evidence in input SMILES tokens to predict their original molecules, leading to less effective training. In addition, in Figure 5, we identify that the optimal value of $\lambda$ is 10, different from the original work on generator-discriminator training in NLP \cite{Clark-ICLR2020} using 50. We suspect that this is because SMILES modeling typically shows smaller losses from the generator than language modeling, and thus we need smaller $\lambda$ to balance the generator and discriminator training.

\paragraph{Architecture analysis.} We analyze diverse variations on the MolTRES architectures, particularly about the architecture of the generator and discriminator. We report ROC-AUC scores on eight MoleculeNet classification tasks and MAE scores on three MoleculeNet regression tasks from each variation. In Table 7, the architecture of our standard setting used in Section 4 is shown in (D). The variations in (A) denote training smaller MolTRES models, showing that reducing layers and hidden size show comparable performance degradation when their numbers of parameters are commensurate. Note that we choose to reduce layers, since it achieves faster model execution speed. The variations in (B) and (C) are about the architecture of generators. (B) contains the variations changing the hidden sizes while using the number of layers of the discriminator, while (C) contains the variations changing the numbers of layers while using the hidden size of the discriminator. In this comparison, we first observe that there is an optimal size of generators that generate training examples suitably challenging for discriminators. After empirical investigation, we choose to set the number of layers in the generator to half of that in the discriminator.

\end{document}